\documentclass[12pt,epsf]{article}
\usepackage{epsfig}

\newcommand{\onefigure}[2]{\begin{figure}[htbp]
\begin{center}\leavevmode\epsfbox{#1.eps}\end{center}\caption{#2\label{#1}}
\end{figure}}
\newcommand{\setfigure}[2]{\begin{figure}[htbp]
\begin{center}\leavevmode\epsfxsize=5in\epsfbox{#1.eps}\end{center}\caption{#2\label{#1}}
\end{figure}}
\newcommand{\twofigures}[3]{\begin{figure}[htdp]
\centering \leavevmode\epsfxsize=2.5in\epsfbox{#1.eps}
\leavevmode\epsfxsize=2.5in\epsfbox{#2.eps} 
\caption{{
#3}\label{#1}}
\end{figure}}

\renewcommand{\thanks}[1]{\footnote{#1}} 

\newcommand{\be}{\begin{equation}}
\newcommand{\ee}{\end{equation}}
\newcommand{\bea}{\begin{eqnarray}}
\newcommand{\eea}{\end{eqnarray}}

\begin{document}

\pagestyle{empty}

\bigskip\bigskip
\begin{center}
{\bf \large Penrose Diagram for a Transient Black Hole}
\end{center}

\begin{center}
James Lindesay\footnote{e-mail address, jlslac@slac.stanford.edu}
and Paul Sheldon\footnote{e-mail address, psheldon@huphysics.howard.edu}
 \\
Computational Physics Laboratory,
Howard University,
Washington, D.C. 20059 \\
\end{center}
\bigskip

\begin{center}
{\bf Abstract}
\end{center}
A Penrose diagram is constructed for a spatially coherent
black hole that smoothly begins an accretion, then
excretes symmetrically as measured
by a distant observer, with the initial and final states described
by a metric of Minkowski form.  Coordinate curves
on the diagram
are computationally derived. Causal relationships
between space-time regions are briefly discussed. 
The life cycle of the black hole demonstrably leaves
asymptotic observers in an unaltered Minkowski
space-time of uniform conformal scale.
\bigskip \bigskip \bigskip

\setcounter{equation}{0}
\section{Introduction}
\indent

Black holes continue to be objects of considerable
analytic and observational interest.  
Because of expected phenomena such as Hawking
radiation and holography, black holes provide meaningful
systems connecting quantum behaviors
to geometrodynamics.  However, that
very coupling to quantum processes implies
that descriptions of the space-time
should qualitatively differ
from classical static systems.  Such changes
in description are needed to model
dynamic space-times
consistent with quantum measurement constraints.

One of the most interesting features of quantum systems
is the space-like coherence of an entangled state.
Motivated by the need to more directly describe such
systems within a geometrically dynamic system,
coordinates describing a spatially coherent
geometry have been developed\cite{JLRiv06, JLNSBP08}. 
The metric form utilizes non-orthogonal coordinates
inspired by the river model of black
holes\cite{rivermodel, Nielsen}. 
Fixed temporal coordinate curves remain space-like
surfaces throughout such dynamic geometries, corresponding
to the times measured by distant, inertial observers.  In addition,
the dynamic horizon and radial mass scale
parameterized using this time have
non-singular curvature and coordinate representations. 
These qualitative differences in the
nearby environment of the dynamic horizon makes
examinations of quantum behaviors more straightforward.

\setcounter{equation}{0}
\section{Form of the Metric and Conformal Coordinates
\label{Section2}}

\subsection{Form of the metric}
\indent

The radially dynamic
space-time metric for a spatially coherent, spherically
symmetric system will be assumed to take the
form\cite{JLRiv06, Nielsen, Clifton}
\be
ds^2 = -\left (1-{R_M (ct,r) \over r} \right ) (dct) ^2 +
2 \sqrt{{R_M (ct,r) \over r}} \, dct \, \, dr + dr^2 + r^2 \, 
(d\theta^2 + sin^2 \theta \, d\phi^2).
\label{metric}
\ee
In this equation, a finite radial mass scale
$R_M (ct,r) \equiv 2 G_N M(ct,r) / c^2$ is the length scale of the
mass-energy content of the geometry, which is given by
the Schwarzschild radius for a static geometry.
The metric takes the
form of a Minkowski space-time
both asymptotically ($r \rightarrow \infty $) as well as when the
radial mass scale vanishes ($R_M(ct,r) \rightarrow 0$).  Therefore,
the temporal and radial coordinates are those of an observer
far from the black hole.  The radial coordinate also provides
the length scale for local tangential distances and areas.
As can be seen from the form of the metric Eqn. \ref{metric},
at the surface instantaneously defined by the radial mass
scale $R_M(ct,r)$, that fixed radial coordinate curve
labeled $r$ transforms from
being time-like external to this scale to being space-like internal to
this scale.  Fixed
temporal coordinate curves are always spacelike, making
generic (non-orthogonal) geometries of this form
convenient for explorations of quantum behaviors. 
This metric has been examined for uniformly accreting\cite{BABJL3}
and excreting\cite{BABJL1,JLQG} black holes in previous articles. 

For a transient black hole initiating at $ct_o$, and
terminating at $ct_f$ the outgoing light-like surface defining
the horizon $R_H$ is given by the null surface of
the metric Eqn. \ref{metric} which vanishes as the decreasing
radial mass scale vanishes at $ct_f$.
This surface satisfies the general equation for
outgoing null geodesics, 
\be
\dot{r}_\gamma = 1 - \sqrt{R_M \over r_\gamma} \quad , \quad
r_\gamma (ct) = 
{R_M (ct,r_\gamma (ct)) \over (1 - \dot{r}_\gamma (ct))^2}.
\label{horizon}
\ee
Several points of interest directly follow from this equation
and the form of the metric:
\begin{itemize}
\item The radial mass scale falls interior to the horizon
while the horizon is expanding $\dot{R}_H>0$, and
exterior to the horizon while it is contracting $\dot{R}_H<0$.
It should be noted that the transition time from mass
accretion to mass excretion does not generally coincide
with that from an expanding to contracting horizon.
\item Since outgoing photons would be momentarily
stationary as they cross the radial mass scale
$\dot{r}_\gamma =0$, there can be no observers with
stationary radial coordinate (fiducial observers) with
$r \le R_M$.  The radial mass scale represents the static
limit for radial motions in this geometry.
\item From the second form of the equation, it is clear that
a finite horizon scale vanishes when the radial mass scale
vanishes, as long as $\dot{R}_H<1$.  Of course, the radial mass
scale \emph{must} vanish if $\dot{r}_\gamma=1$.
\item  There are trapped outgoing light-like surfaces which
initially propagate with
$\dot{r}_\gamma >0$ through
non-vanishing radial mass scale $R_M$ 
with radial coordinates that vanish as
$ct \rightarrow ct_o$, as well as at some final
time $ct < ct_f$ on the singularity. 
These trajectories will be demonstrated later
in Figure \ref{Fig3a}b.  This means that
$R_M / r_\gamma < 1$ for times infinitesimally
close to $ct_o$.
\item Since distinct outgoing light-like trajectories
can initiate on the surface $(ct=ct_o,r_\gamma=0)$,
for spatially coherent dynamic geometries
the initiation of the (possible) singularity
should be represented by an ingoing
light-like surface. 
The curve $r=0$ should transform through this surface
from time-like to space-like.
\end{itemize}

Radial trajectories in this geometry have 4-velocity
components that satisfy
\be
u^r =-u^0 \sqrt{R_M \over r} \pm \sqrt{(u^0)^2 - \Theta_m}
\quad , \quad \Theta_m \equiv \left \{
\begin{array}{l}
1 \quad m \not= 0 \\
0 \quad m = 0
\end{array}  \right .
,
\ee
where the + sign signifies outgoing trajectories, and the
- sign signifies ingoing trajectories.  It should be noted
that for massive systems, trajectories $u^0=1$
are neither ingoing nor outgoing, and they represent what
have been referred to as \emph{geometrically stationary}
trajectories\cite{JLQG}.  Trajectories with 4-velocity
components satisfying
$u^0=1, \, u^r=-\sqrt{R_M \over r}$
satisfy geodesic equations for massive gravitating
systems which share proper time with asymptotic observers,
$dt=d\tau$.

The static form of this geometry
$\dot{R}_M =0$ satisfies the same
form of dominant energy conditions as does a generalization
of orthogonal Schwarzschild space-time with a 
spherically symmetric static mass distribution.  
A spatially coherent system whose
temporal variations are sufficiently slow can likewise satisfy
these conditions.  One can therefore directly construct
models of physical
systems whose only violations of energy conditions are
consistent with those due to quantum effects.

Since the large scale causal structure
of the space-time represented by the Penrose diagram
should not depend sensitively
on the form of the radial mass scale, one
can choose a relatively simple form for its functional
dependence.
For the present calculation, the radial
mass scale will be taken as a time-dependent
form given by the ``bump function"   
\be
R_M(ct) \equiv R_{max} \, exp \left ( {(ct_f)^2 \over (ct)^2 - (ct_f)^2}
 \right ) 
\ee
during the period $ct_o=-ct_f \le ct \le ct_f$. 
\twofigures{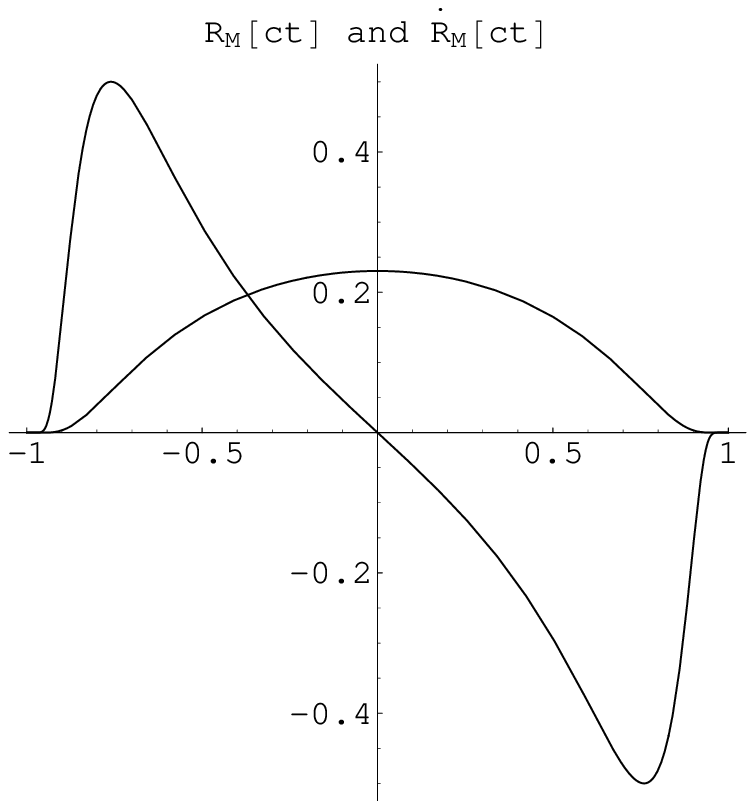}{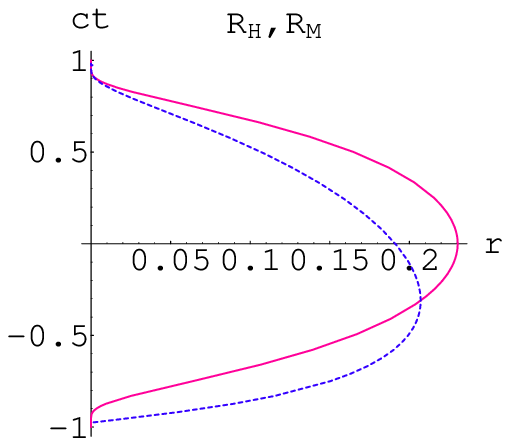}{Left: Bump function and its derivative.
Right: Horizon (dotted, blue) and
radial mass scale (solid, red). }
All curvature components generated
by this metric are non-singular away from the physical singularity
$r=0$ as seen from the Ricci scalar
\be
\mathcal{R}=3 \left ( {ct_f \over (ct)^2 - (ct_f)^2} \right )^2 
{ct \over r}
\sqrt{R_M (ct) \over r} .
\ee
This means that no observer measures singular curvatures
at the horizon, radial mass scale, or transition times
of the geometry.

\subsection{Development of conformal coordinates}
\indent

The technique used to generate the conformal diagram
relied only upon constructing light-like surfaces for the
metric Eqn. \ref{metric}.  The conformal coordinates
will be labeled $(v,u)$.  For ingoing null geodesics,
the required equation takes the form
\be
\dot{r}_v = -1 - \sqrt{R_M \over r_v} \, ,
\label{rvEqn}
\ee
while outgoing null geodesics satisfy
\be
\dot{r}_u = 1 - \sqrt{R_M \over r_u} \, .
\label{ruEqn}
\ee
Ingoing light-like trajectories labeled by
$v$ can access all regions
of space-time through a past Minkowski space-time
correspondence, initiating on past light-like
infinity $skri-$.  For the chosen geometry, 
ingoing light-like trajectories are demonstrated
in Figure \ref{Fig2}.
\onefigure{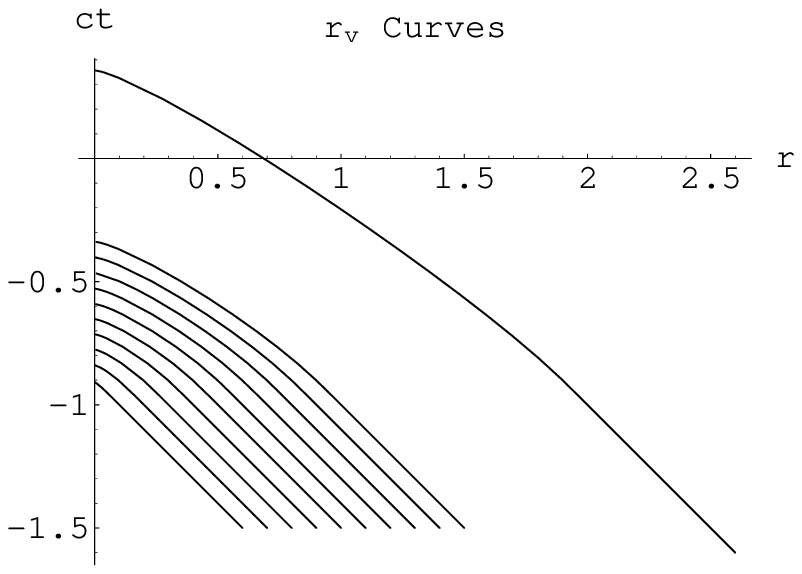}{Ingoing light-like trajectories, labeled
by conformal coordinate $v$.}

Likewise, for the region exterior
to the horizon, outgoing light-like trajectories
labeled by $u$
passing through any point have a future
Minkowski correspondence, terminating on future
light-like infinity $skri+$.
One only needs to develop labels $u$
for outgoing light-like trajectories
in the interior of the horizon.
The functional representation of the singularity
on the conformal diagram $u=u_{r=0} (v)$
defines an analytic extension that parameterizes these 
trajectories in terms of their termination values $v$
on the singularity.  Several
outgoing light-like trajectories for the given geometry
are demonstrated in Figure \ref{Fig3a}.
\twofigures{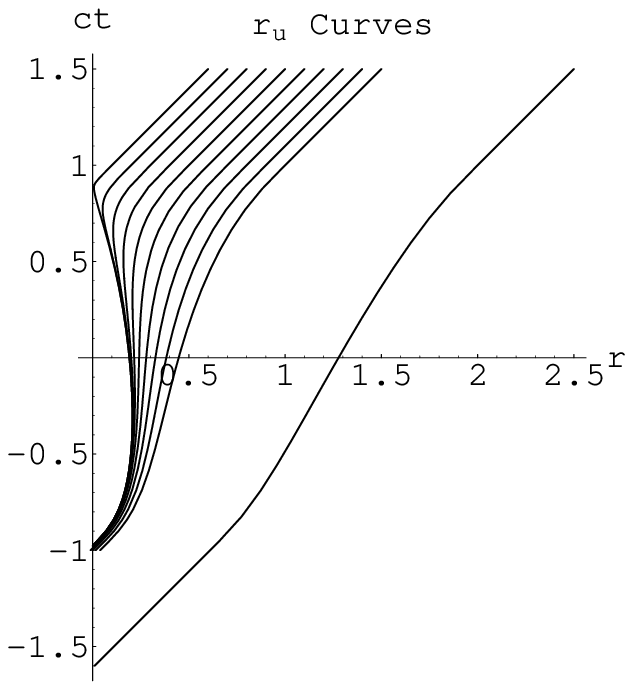}{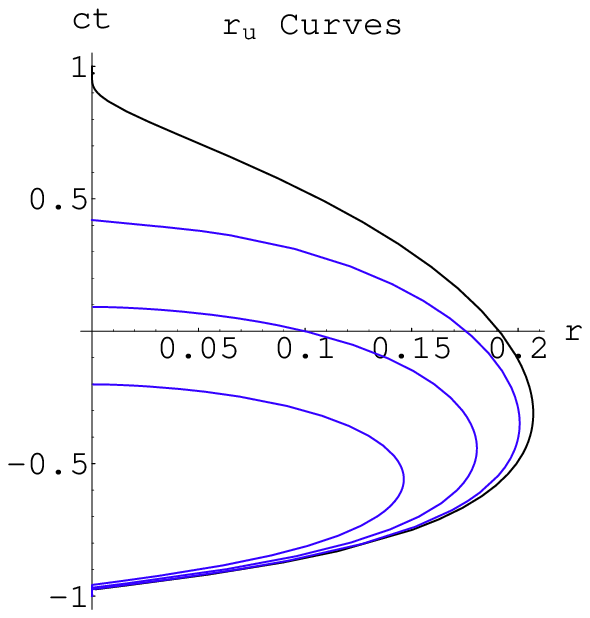}{Exterior (a) and interior (b)
outgoing light-like trajectories, each labeled by
the value of conformal coordinate $u$.}
The trajectories of outgoing photons near the
horizon are of particular interest.  An observer with
a fixed radial coordinate external to the radial mass
scale and horizon will not observe photons emitted
near the horizon until after the singularity has vanished. 
This means that an object falling through the horizon
will not be observed to cross the horizon until after
the black hole is no longer present.  The moment
just prior to the observation of the final evaporation
of the singularity is directly seen to provide all observable
information on matter that fell through the horizon
during its period of transience.  Coherence relationships
with exterior constituents need not be disrupted as
infalling constituents cross the horizon.

The curves representing the radial centers $r=0$
of the initial and final state
Minskowski space-times of the transient black hole
should not be expected to be co-linear vertical lines. 
Since light-like trajectories get shifted relative to those
in flat space-time due to gravitational attraction, 
the conformal Minkowski
coordinates $(v=ct+r+v_s, u=ct-r+u_s)$ are relatively shifted
by factors $(v_s,u_s)$ that are fixed on the
particular trajectory.  For the present calculations,
the label $v$ will be taken from the past Minkowski initial
state surface $v=ct_o + r_o$, while the exterior labels $u$
will be taken at the future Minkowski final state surface
$u=ct_f - r_f$, providing boundary conditions for the
differential equations \ref{rvEqn} and \ref{ruEqn}.  These labels
can be directly observed from the surface
$ct_o=-1$ on Fig. \ref{Fig2} and $ct_f=+1$ on
Fig. \ref{Fig3a}a.

\setcounter{equation}{0}
\section{Penrose Diagram of the Transient Black Hole
\label{Section3}}
\indent

Space-time diagrams can be quite useful for
visualizing the dynamic relationships in a given
geometry.  
Penrose diagrams are
convenient for examining the large-scale causal
structure of a geometry because of the following
properties:
\begin{list}{...}{}
\item Penrose diagrams map the entire space-time
onto a single finite page, and
\item Penrose diagrams are conformal diagrams,
i.e., they preserve the slope of
light-like trajectories at $\pm$unity.
\end{list}
Light-like surfaces define the boundaries of causal
regions in space-time,  so
the causal structure of the geometry
can be directly observed from any conformal diagram, and
potential causal relationships between events
can be immediately ascertained.  
Minkowski coordinates $(ct,r)$ of flat space-time are already
conformal coordinates.  
However, while $R_M \not= 0$ for the metric \ref{metric}, 
null geodesics do not have unity slope for
coordinates $(ct,r)$ as seen in Figures \ref{Fig2}
and \ref{Fig3a}.  Therefore, the light-cone conformal
coordinates $(v,u)$ will be used to construct
the Penrose diagram for the transient black hole.

\subsection{Procedure for constructing the Penrose diagram}
\indent

The conformal diagrams presented use hyperbolic tangents of a
scaled multiple of the conformal coordinates $(v,u)$ 
to map the infinite domain of those conformal coordinates onto a
finite region.  
The coordinate $Y_{\uparrow}$ labeling the
vertical axis takes the
form $(tanh({v \over scale}) + tanh({u
\over scale}) ) / 2$ and the coordinate $Y_{\rightarrow}$
labeling the horizontal axis takes the
form $(tanh({v \over scale}) - tanh({u \over scale}) )/ 2$, so
the diagram has its domain and range bounded by $\pm 1$. 
In the interior region $r<R_H (ct)$, 
 the analytic continuation
must be chosen to generate a representation for
the singularity that is space-like (except perhaps at
the endpoints). 
Since the entire coordinate system has been generated
using light-like trajectories
from Eqns. \ref{ruEqn} and \ref{rvEqn},
the slopes of \emph{any} outgoing/ingoing
light-like radial trajectories on the diagram
will automatically be $\pm 1$. 
The overall causal structure of the transient black
hole is demonstrated in Figure \ref{Fig4}.
\onefigure{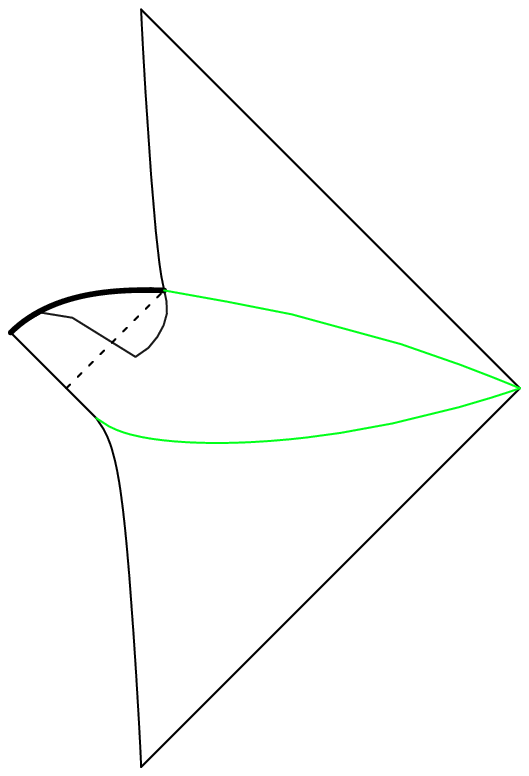}{Horizon, radial mass scale,
singularity, and overall
causal structure of the black hole.}
The figure demonstrates the initial and final time volumes
as green horizontal curves terminating at the right corner
of the diagram.  Prior to the initial volume state
and subsequent to the final volume state, the space-time
is that of Minkowski.  The diagram is bounded
on the right by past light-like infinity $skri \, minus$
(bottom right) and future light-like infinity $skri \, plus$
(top right).  

The diagram is bounded on the left by
the center $r=0$, which is space-like during the
period of transience of the black hole, light-like
during the initiation of the singularity, and otherwise
time-like.  The light-like surface initiating the singularity
has coordinates $(ct=-1 unit, r=0)$. 
This ingoing light-like surface expands the
conformal space-time\cite{BABJL3} in a manner
differing from standard textbook treatments\cite{LSJLBH,Birrell}.
The horizon is the dotted surface with
unit positive slope, which is crossed once by the solid
curve representing the radial
mass scale.  The interior region is left of the horizon, bounded
from above by the singularity.  A more detailed description
will be given in the next section.

The initial and final geometries are
Minkowski space-times for $|ct| \geq 1 unit $.  The bump function
smoothly transitions the dynamic metric
in Eq. \ref{metric} into static forms.   
The radial coordinates must smoothly match across the
transition volumes $|ct|=1 unit$, since $4 \pi r^2$ measures the area
of any sphere of radial coordinate $r$ for both
local conformal coordinates
$(v,u,\theta,\phi)$ as well as those of the asymptotic observer
$(ct,r,\theta,\phi)$.  

\subsection{Features of the Penrose diagram}
\indent

The Penrose diagram in Figure \ref{CoordPenrose} demonstrates
the expected global structure of this spherically symmetric,
spatially coherent black hole that
smoothly varies the radial mass
scale $R_M (ct)$ with respect to the
distant observer's time coordinate $ct$.  
\setfigure{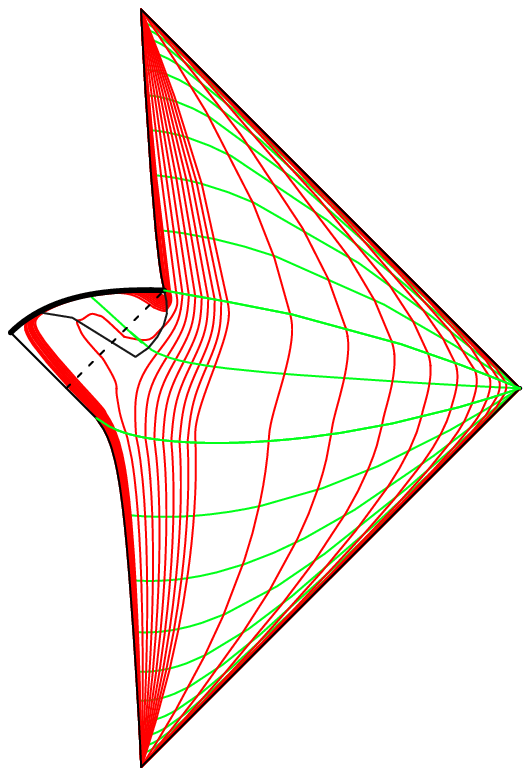}{Penrose diagram for a black hole that 
symmetrically accretes from zero mass at $ct=-1$ then excretes
to zero mass at $ct=+1$.  Red curves (running vertically in the 
exterior right
hand region) represent curves of constant $r$.  The green curves
represent curves of constant $ct$.  The dashed line
represents the horizon, and the solid, bold space-like
curve represents the singularity. \label{CoordPenrose}}
The parameters were chosen such that the maximum rate
of mass accretion/excretion is $|\dot{R}_M | \le 0.5$,
yielding a maximum
radial mass scale $R_M (0) \cong 0.23 units$, which gives a
maximum horizon scale of $R_H (-0.31 units) \cong 0.21 units$.
In the diagram, the red curves that are time-like
(vertical) in the right hand
regions represent curves of constant $r$, originally graded from
$r=0$ in hundredths, tenths, then in units of the chosen scale. 
The curves of constant radial coordinate $r$ all originate at the bottom corner of the
diagram representing $t=-\infty$, and terminate at the uppermost corner
representing $t=+\infty $.  The green curves, which,
away from initiation or termination surfaces, are 
everywhere space-like (horizontal), 
represent curves of constant $ct$ graded in units
of the given scale.  All constant $ct$ curves originate on the curve $r=0$
and terminate at the far right corner of the diagram representing $r=\infty $. 
The various $ct=constant$ and $r=constant$ curves each intersect at only
one point on the diagram (except for the
initiation of the singularity, and light-like infinities). 
The light-like bounding curves $r=\infty$
on the right are those of a fixed scale Minkowski space-time.

One should note that at $ct=-1 unit$, the singularity develops via
an ingoing light-like transition $(ct=ct_o, r=0)$, expanding
the available conformal space-time.  
This new region, represented as the expanded area
in the middle left portion of Figure \ref{CoordPenrose},
contains most of the significant features
of the black hole.
The curve $r=0$, which bounds the diagram on the left,
is initially a time-like trajectory bounding the initial Minkowski
space-time.   However, due to relative parameter shifts 
because of the transient black hole, this curve is not 
represented by a vertical line on this diagram. 
Rather, this curve smoothly joins the curve $ct=ct_o$
at the transition. 
As the singularity forms, the
curve $r=0$ undergoes a light-like transition
to become a space-like trajectory bounding the upper
interior region of the black hole from above.  As previously
mentioned, the form of this space-like surface
and the transition to it \emph{defines}
the analytic continuation of correspondence with the
exterior coordinates.  As the singularity vanishes,
the curve $r=0$ bounds the final Minkowski space-time
from the left, again skewed from the vertical. 
The radial mass scale is indicated by
the solid dark gray curve initiating at the far left
corner of the diagram, crossing
the horizon $R_H$ near $t=0$, then terminating as
the singularity vanishes. 
As can be seen from Eqn. {\ref{horizon}}
the radial mass scale $R_M (ct)$ 
lies within the horizon during growth of the horizon and outside of
the horizon during shrinking of the horizon\cite{JLRiv06}. 
Radial coordinate
curves are seen to transition from time-like to space-like as
they cross the radial mass scale (as
expected from the metric
Eqn. \ref{metric} for fixed radial coordinate $dr=0$).  
As previously noted,
outgoing light-like trajectories Eqn. \ref{ruEqn}
are momentarily stationary in the radial
coordinate $\dot{r}_u = 0$ at the radial mass scale.

The dynamic horizon is represented
by the dashed diagonal line 
in Figure \ref{CoordPenrose}, terminating as the
space-time reaches its final Minkowski form.
It is noteworthy that the horizon
lies completely within the expanded region of
space-time to the left of this Minkowski geometry. 
The black hole horizon has a finite temporal duration;
unlike the case for a static Schwarzschild black hole, 
both temporal and radial
coordinate curves are seen to cross this surface, i.e.
the horizon $R_H (ct)$ is \emph{not} a $t=\infty $ surface.

\subsection{Causal regions}
\indent

Penrose diagrams are most convenient for exploring the
large-scale causal structure of the space-time.  For the
transient black hole, various causal regions, are represented
in Figure \ref{Fig6}.
\setfigure{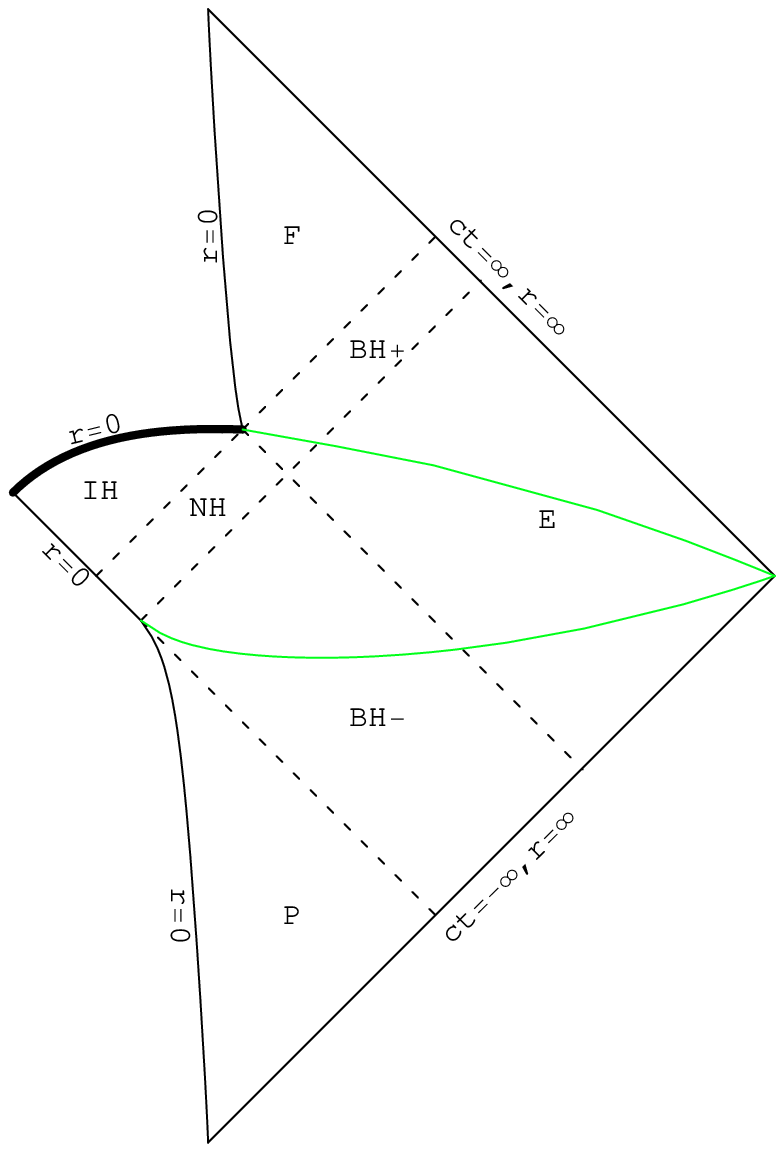}{Causal regions include interior to the horizon IH,
near horizon NH,  exterior E, causal past BH-, causal future BH+,
past P, and future F.}
The boundaries of the regions are shown as dotted
light-like surfaces on the Penrose diagram. 
It is of considerable interest to examine causal relationships with
the region interior to the horizon labeled IH.  In particular, one
might note that events in this region can share no direct
causal relationship with events in
the future region F (i.e., events in these regions share neither
time-like, light-like, or space-like relationships). 
This means that even a
quantum entangled relationship cannot be established
between disparate events within these regions. 
However, a space-like
correlated event associated with the interior
can be a cause to an event in future region F. 
The interior region \emph{can}
have space-like entanglements
(coherence) with systems in the near horizon
(NH), exterior (E), black hole causal past (BH-), and
black hole causal future (BH+) regions. 
Boundary conditions
for any quantum systems with space-like coherence with the
interior should be physically consistent
across these regions of the space-time.  Explicitly, interior
entanglements cannot have boundary conditions within
the future region F.

Other causal relationships are indicated in Fig. \ref{Fig7}.
\setfigure{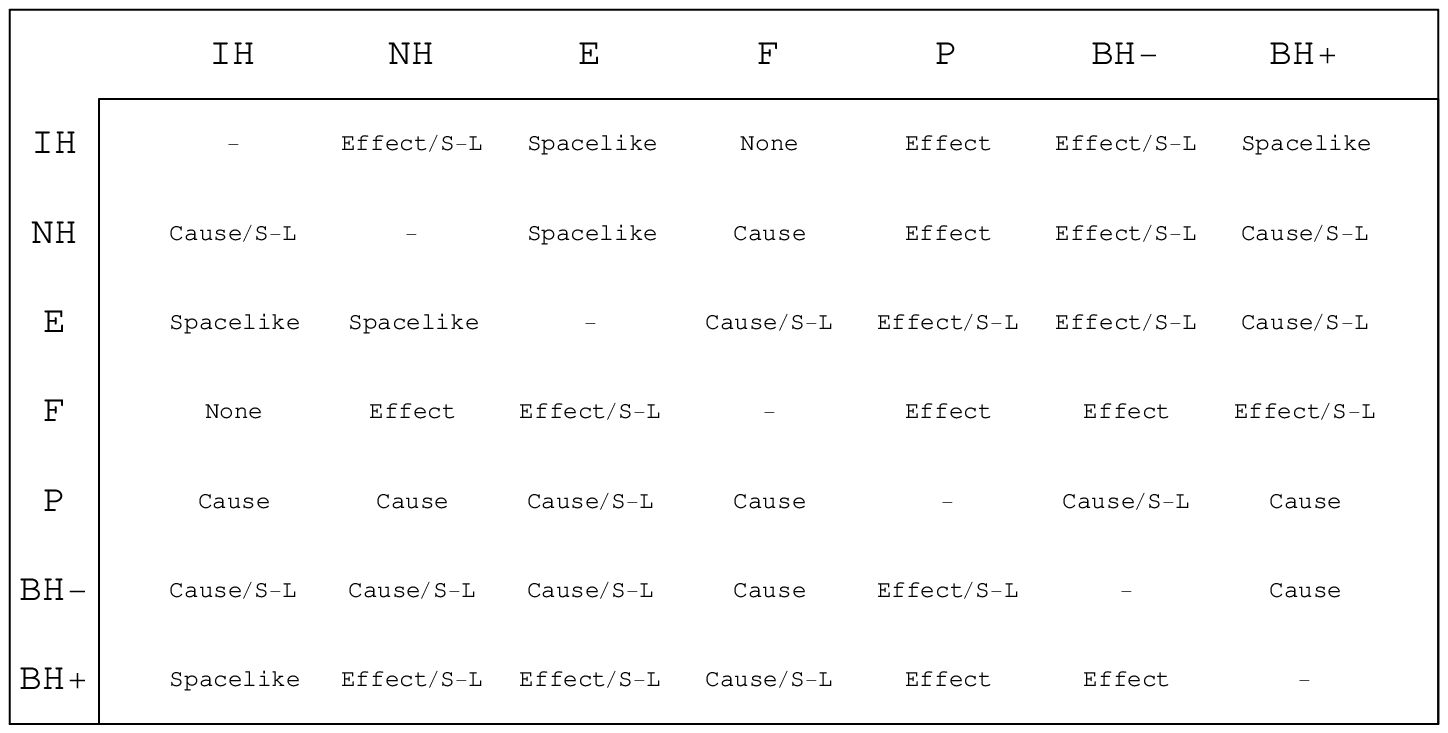}{Causal relations.  Events in a region
labeled by a given row 
possibly relate to those in a region labeled
by a given column as indicated.}
The possible relationships are indicated as either
cause, effect, space-like (S-L), or none.  Cause/effect
relationships must be either (past/future) time-like or
light-like. 
These relationships are completely consistent with
those reported earlier in reference \cite{BABJLNSBP09}.

\setcounter{equation}{0}
\section{Conclusions and Discussion}
\indent

Generic spherically symmetric, dynamic geometries
with mass scale $M=M(ct,r)$ can be used to model
dynamic physical systems.  A spatially
coherent dynamic black hole has been shown to
maintain space-like, fixed time volumes throughout
the global geometry, while generating a space-like
surface $r=0$ and an outgoing light-like horizon
during its period of generating significant curvature. 
Parametric descriptions of both the radial mass scale
and the horizon are non-singular in the coordinates $(ct,r)$, 
as are scalar forms of physical curvatures near
these trajectories.  The radial mass scale does
not coincide with the light-like horizon.

The large-scale geometry of a transient black hole has been seen to
have negligible effect upon the scales of distant observers. 
The past and future light-like infinities of the black hole are the
same as those of Minkowski space-time.    
The onset of the singularity and a finite mass scale
expands the previously low-curvature conformal space through
an ingoing light-like transition, 
concurrently expanding
the large-scale structure of the space-time.

\begin{center}
\textbf{Acknowledgments}
\end{center}
The authors would like to dedicate this article to the late
Dr. Beth A. Brown of NASA Goddard, who initiated the
research program which culminates in this paper.  JL 
wishes to acknowledge the intellectual inspirations
of Lenny Susskind, as well as being introduced to
non-orthogonal coordinates by James Bjorken.  He
also would like to recognize useful discussions
with Tehani Finch.
PS would like to thank Prof. Wolfgang Rindler for
having patience in his long studies of time asymmetry,
and Prof. H. Dieter Zeh for giving him courage
in his studies of the problem of time. 
He also acknowledges
Professors Marlan Scully, Yuri Rostentov, and
Andre Matsko for scholastic encouragements.

\end{document}